# Atomistic simulation of light-induced changes in hydrogenated amorphous silicon

T. A. Abtew[1] and D. A. Drabold[1]

**Abstract:** We employ *ab initio* molecular dynamics to simulate the response of hydrogenated amorphous silicon to light exposure (Staebler-Wronski effect). We obtain improved microscopic understanding of PV operation, compute the motion of H atoms, and modes of light-induced degradation of photovoltaics. We clarify existing models of light-induced change in aSi:H and show that the "Hydrogen collision model" of Branz[3] is correct in essentials.

Solar photovoltaic (PV) devices are an increasingly important source of electrical power. Current PV production is entering the giga-watt regime, so that a factor of the order $10^3$ is needed in order to impact the (terawatt scale) energy markets[1]. One material used for PV energy production is hydrogenated amorphous silicon (a-Si:H), which is particularly inexpensive and serves an important niche in energy markets. An impediment to the use of a-Si:H cells is the Staebler-Wronski effect[2] (the light-induced creation of carrier traps causing reduced energy conversion efficiency). A salient feature of a-Si:H is that exposure to intense light (as in PV applications) leads to structural change (rearrangements of the positions of the atoms in the amorphous network). These changes have a serious impact on the performance of a-Si:H cells (PV efficiency drops 15% - 20%), and then stabilizes. Such light-induced structural changes [named the Staebler-Wronski Effect, (SWE)] are complex and information about the changes is provided through an array of experiments[4,5]. Key experimental facts are: (1) Light-soaking induces large changes in photoconductivity, and defect (carrier trap) formation; (2) light induces H motion[3,6] (3) light soaking preferentially creates protons separated by 2.3 Å in device grade material and a shorter distance in low quality material[7], (4) Isoya and coworkers[8] showed that no dangling bond (DB) pairs are formed after light soaking and that the placement of the light induced dangling bonds was random[9]. Further, it was shown that metastable dangling bonds were separated by at least 10 Å, (5) studies of defect creation and annealing kinetics in a-Si/Ge:H suggest that there is not a large population of mobile H leading to recapture events of H onto DBs as part of the photo-degradation process[10].

The most widely accepted model of photo-degradation is the "Hydrogen Collision Model" of Branz[11]. Here, recombination-induced emission of H from Si-H bonds creates both mobile H and vestigial DB's. When two such mobile H atoms join in a metastable Si-H complex, two new DBs become stable. Our work lends support to this view.

[1] Department of Physics and Astronomy, Ohio University, Athens, OH 45701, USA.

Associated with Branz's work is the two-phase model of Zafar and Schiff, which successfully explained thermal metastability data[12,13], exploited the concept of paired H, and was later merged with the model of Branz and invoked dihydride bonding[14], which emerges naturally in our simulations. A seminal paper emphasized local "weak bond" models[15], which have been ruled out by the non-locality of rearrangements inferred from Electron Spin Resonance[8]. Recently we combined experimental information with the results of accurate simulation in high quality a-Si:H to show that a likely consequence of light-soaking is the formation of Si sites bonded to two network Si atoms and two H atoms (we label this class of structures $SiH_2$)[16].

A predictive simulation of these effects must include several ingredients: (1) models of a-Si:H representative of the topology of the network, sufficiently large to faithfully represent short-time (*ca* several picoseconds) dynamics, (2) accurate interatomic interactions (the H energetics are highly delicate in a-Si:H[17,18]), (3) the electron-lattice interaction must be estimated in a reasonable fashion[4,19] (4) the motion of H must be included, as a variety of experiments point to the role of H motion in the SWE. We accommodate all of these requirements in our work, which sets this study apart from other simulations of these effects.

Generating a representative plausible model is the first step in incorporating the ingredients stated above. Our starting point in getting this model is a 64-atom defect free (four-coordinated) a-Si model[20]. To create the hydrogenated amorphous silicon environment, we removed three silicon atoms and added 10 more hydrogen atoms to terminate all but two dangling bonds. This procedure generates a 71 atom hydrogenated amorphous silicon model. We relaxed this starting configuration using conjugate gradient for coordinate optimization. We then repeated this supercell surgery at other sites in order to generate an ensemble of models (3 configurations).

Topological or chemical irregularities in an amorphous network lead to localized electron states in the gap or band tails[21]. If such a system is exposed to band gap light, it becomes possible for the light to induce transitions of electrons from the occupied states to low-lying unoccupied (conduction) states. For the present calculation we do not concern ourselves with the subtleties as to how the EM field induces such transitions; we will simply assume that a photo-induced promotion occurs, by depleting the occupied states of one electron "forming a hole" and moving the electron near the bottom of the unoccupied "conduction" states. Changes in force due to the light-induced transition of carriers will initially be local to the region in which the orbitals are localized, followed by transport of the thermal energy through the network. In general, it is necessary to investigate photo-structural changes owing to a collection of different initial and final states, though we may expect that only well localized states (necessarily near the gap) have the potential to induce structural change[22,23,24,25].

To begin unraveling the SWE, we track the dynamics of a-Si:H in the presence of electron-hole pairs. We have performed extensive molecular dynamics simulations of network dynamics of a-Si:H both in the electronic ground state and also in a photo-excited state (in the presence of a simulated electron-hole pair) using our model. The density functional calculations were performed within the generalized gradient approximation (GGA) using the first principle code SIESTA[26] at a constant temperature (300K) We have used a fully self-consistent Kohn-Sham functional in the calculation of energies and forces within the parametrization of Perdew, Burke and Ernzerhof[27]. Norm conserving Troullier-Martins[28] pseudopotentials factorized in the Kleinman-Bylander[29] form were used. We employed an optimized double- polarized basis sets (DZP), where two *s* and three *p* orbitals for the H valence electron and two *s*, six *p* and five *d* orbitals for Si valence electrons were used. We found that to obtain the right geometry and dynamics for hydrogen, these high quality approximations were required[18].

We computed the atomic diffusion in a light-excited state and the electronic ground state, by computing the mean-square displacement (MSD) of Si and H atoms. We use a definition, $MSD = \frac{1}{N_\alpha} \sum_{i=1}^{N_\alpha} |r_i(t) - r_i(0)|^2$ , where the sum is over particular atomic species $N_\alpha$ and $r_i(t)$ are the coordinates of the atom at time *t*. In Fig. 1, we plot the mean-square displacement, averaged over all H atoms and Si atoms, respectively, in the network, for both the electronic ground state and a light-excited state. Consistent with the measurements of Isoya[8], the hopping of H is strongly stimulated by the electron-hole pairs. The enhanced diffusive motion of H in the photo-excited state relative to the electronic ground state arises from the strong electron-lattice interaction of the amorphous network, and an effect of "local heating" in the spatial volume in which the state is localized.

By performing proton NMR measurements with the stable $SiH_2$ site in a lower quality (higher defect density) material, Bobela *et al*[30] extracted the distribution of proton-proton distances in the material, producing a quite Gaussian curve with a mean of 1.8Å and a FWHM of 0.54Å. In Fig. 2, we show the results for a room temperature thermal simulation, where we track the H-H distances through the course of a 250 fs, 300K molecular dynamics run. There is reasonable agreement between the proton NMR data and our thermal simulation. For the "lower quality" material, the two "short-distance" conformations indicated are more important than for the better material.

In addition to the enhanced hydrogen diffusion, we have observed another important difference in the two cases. In the light-excited state, we observed a preferential formation of $SiH_2$ structures with an average H-H distance of 2.37 Å. The formation of this new structure ($SiH_2$) follows the dissociation of H atom from a Si-H bond close to the dangling bond, which suffered an occupation change, and followed by diffusion through interstitial sites. The H atom then diffuses until it becomes trapped at a variety of centers such as dangling bond (Si-DB), Si bond center, or other defect site. This mobile hydrogen atom is not only trapped at a site but also breaks a Si-Si bond to form another Si-H bond and by doing so, it introduces additional dangling bond(s) in the system. The additional dangling bonds, arising from the enhanced diffusion of hydrogen, are one of the features observed in SWE. These processes continue until two diffusing hydrogen atoms form a bond to a single Si atom to form a metastable $SiH_2$ conformation. We have repeated this simulation for three different configurations and in every light-excited case a paired-H final state was obtained. However, consistent paired H formation, was *not* observed for thermal MD simulation in the ground state. A representative result is given in Fig. 3, which illustrates the motion of the H atoms in the photo-excited state for the aSiH-71 model and shows the formation of $SiH_2$ via the pathway indicated. The snapshots[30] are taken for intermediate steps in the dynamics of the whole cell, in the simulated light-excited state, where emphasis is given to the two hydrogen atoms, which eventually be part of the $SiH_2$ conformation. This strongly confirms a basic event of the H collision model[3] and other pairing models[14].

Our conclusion is that highly accurate simulations in a light-induced state lead to the formation of a class of new configurations, consistent (a) with recent NMR experiments[7,30] and our previous studies[16], and (b) with the hydrogen collision model of Branz[3] and other paired hydrogen model[14]. In contrast, simulations in the electronic ground state do *not* exhibit the tendency to paired-H final states. For the first time, we show the detailed dynamic pathways that arise from light-induced occupation changes, and provide one explicit example (Fig. 3) of defect creation and paired H formation. The utility of suitable *ab initio* methods for modeling such light-induced processes has been demonstrated. This work should be extended in a number of ways, among them: exploration of longer time scales, large models, and a thorough study of possible photo-excitations and the statistics of events associated with them.

**Acknowledgements** NSF-DMR 0310933, 0205858 supported this work. We thank P. C. Taylor, E. A. Schiff, and P. A. Fedders for very helpful discussions and N. Mousseau for models of a-Si:H.

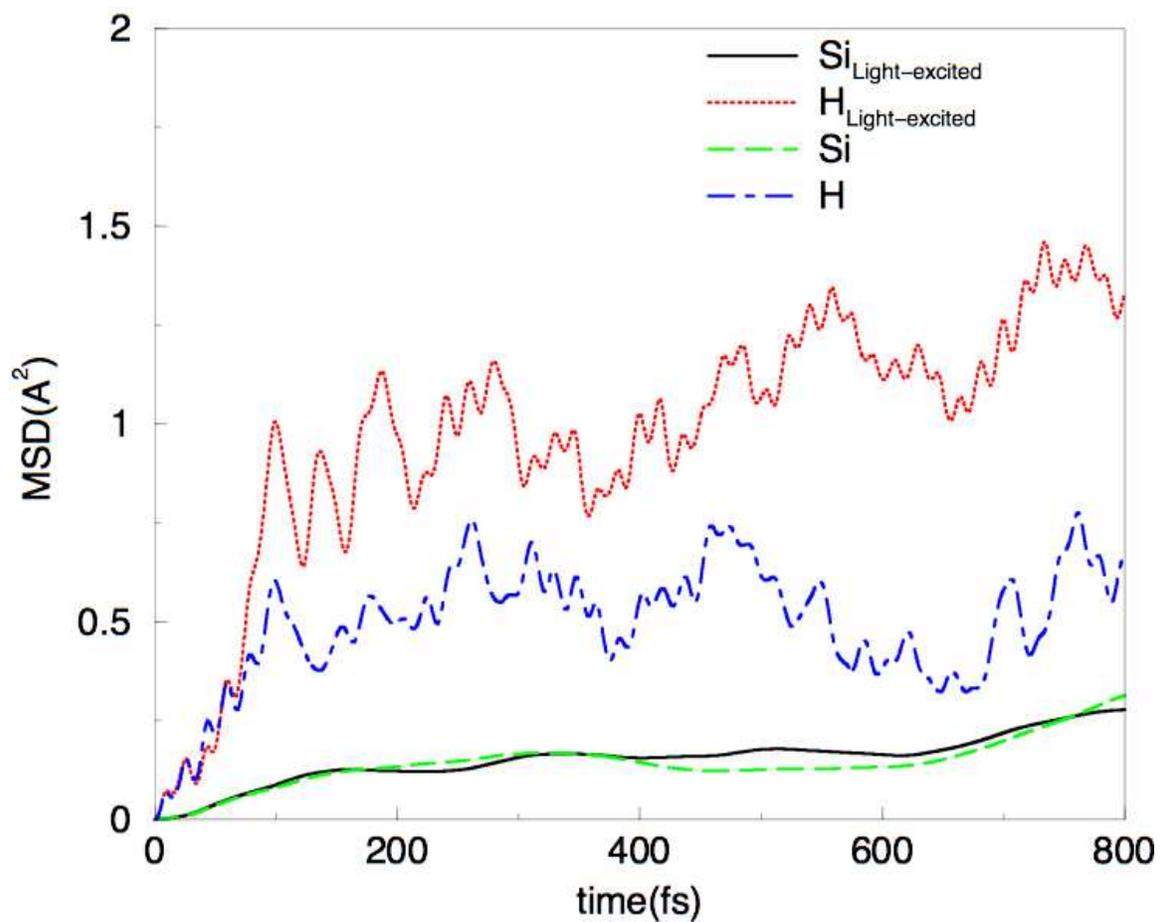

FIG. 1: (Colour online) Mean square displacement (MSD) for silicon and hydrogen atoms at T=300K in the ground state (Si and H) and light-excited state ($Si_{light-excited}$ and $H_{light-excited}$).

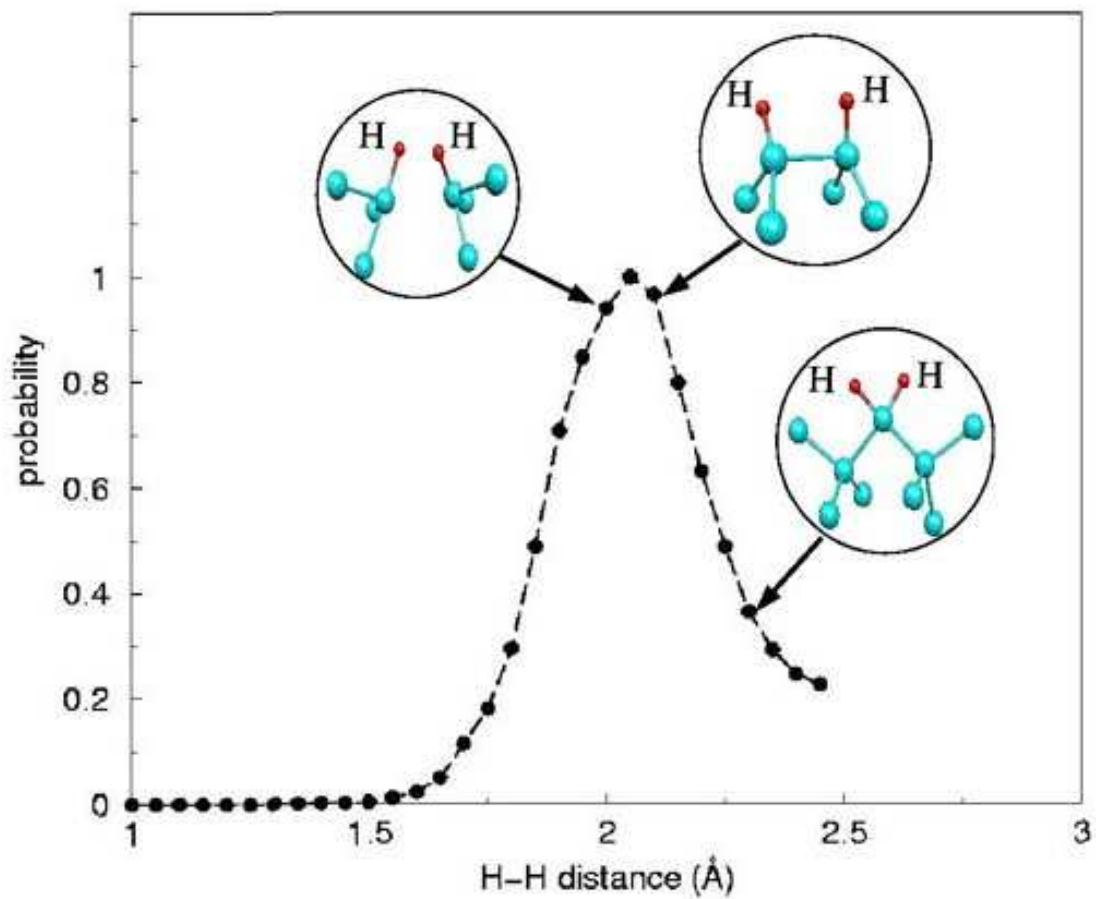

FIG. 2: (Colour online) Distribution of H-H distances from 300K constant temperature molecular dynamics simulation. Insets: configurations primarily responsible for weight near indicated separations. Mean is 2.08Å, FWHM 0.60

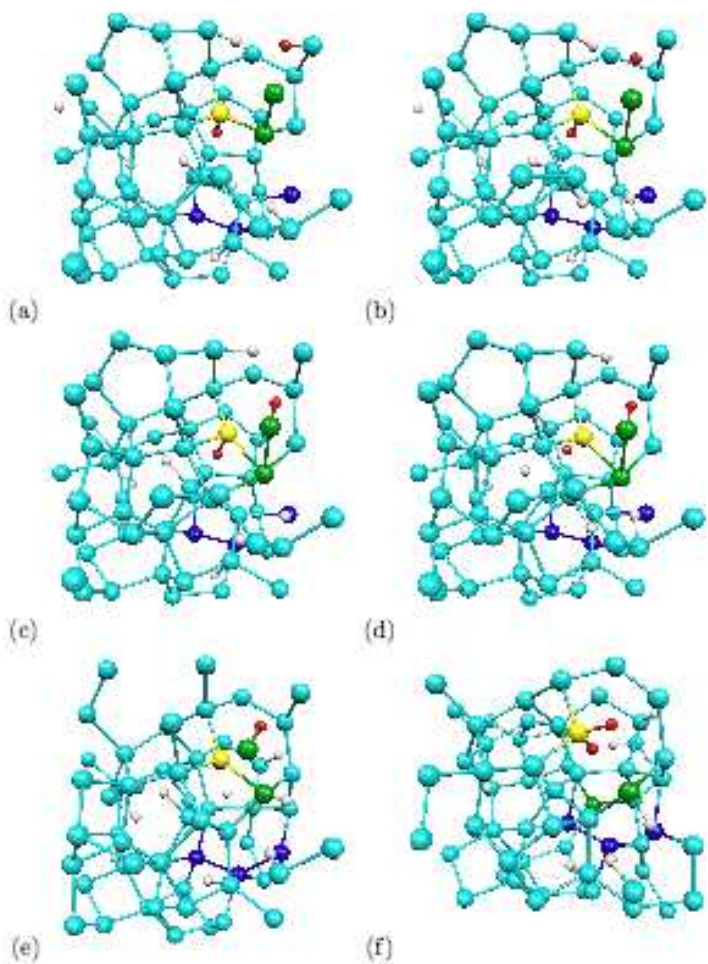

FIG. 3: Motion of key atoms from photo-excited *ab-initio* MD simulation of a-Si:H. Color-coding: aqua: "typical" Si atoms, white, "typical" H atoms, red: H atoms that eventually pair, green: initial Si DB sites, blue: Si atoms that end as DB. a) Initial conformation, b) One H dissociates and forms DB, c) Mobile H attaches to a Si DB, d) The other red H becomes mobile, e) Atomic rearrangements near defect sites, f) $SiH_2$ is formed.